\documentclass{article}

\usepackage{spconf,amsmath,graphicx,hyperref}
\usepackage{xcolor}
\usepackage{amsfonts}
\usepackage{booktabs}
\usepackage{multirow}
\usepackage{amssymb}
\usepackage{graphbox}
\usepackage{xspace}
\hypersetup{hidelinks}
\usepackage{capt-of}
\usepackage{graphicx}
\usepackage{xurl}
\usepackage{tikz}
\usepackage{siunitx}
\usetikzlibrary{arrows.meta, shapes.geometric, calc, math}

\def\x{{\mathbf x}}

\newcommand{\sparc}[0]{\text{SAB}\xspace}
\newcommand{\base}[0]{\text{UAB}\xspace}
\newcommand{\synth}[0]{\text{syn}}
\newcommand{\dys}[0]{\text{nat}}

\newcommand{\pt}[0]{\mathtt{PT}}
\newcommand{\ft}[0]{\mathtt{FT}}

\newcommand{\Dsynth}[0]{$\mathcal{D}_{\synth}$\xspace}
\newcommand{\Dsynthclean}[0]{$\mathcal{D}^{\text{(clean)}}_{\synth}$\xspace}
\newcommand{\Dsynthdys}[0]{$\mathcal{D}^{\text{(dys)}}_{\synth}$\xspace}
\newcommand{\Duaspeech}[0]{$\mathcal{D}_{\dys}$\xspace}

\definecolor{orcidgreen}{HTML}{A6CE39}

\DeclareRobustCommand{\orcidlogo}{%
  \begin{tikzpicture}[baseline=-0.6ex]
    \fill[orcidgreen] (0,0) circle[radius=0.16];
    \node[white] at (0,0) {%
      \fontfamily{qag}\selectfont\upshape\tiny ID};
    \fill[white] (-0.0625,0.095) circle[radius=0.007];
  \end{tikzpicture}%
}

\DeclareRobustCommand{\orcid}[1]{%
  \href{https://orcid.org/#1}{\orcidlogo}%
}

\title{Enhancing Pathological Speech through Articulatory Bottlenecks}
\name{Ailín Pollio San Pedro\orcid{0009-0009-9778-3104}, Olivier Perrotin\orcid{0000-0002-9909-6078}, Thomas Hueber\orcid{0000-0002-8296-5177}}
\address{University Grenoble Alpes, CNRS, Grenoble INP, GIPSA-lab, France}
\begin{document}
\ninept
\maketitle
%

\begin{abstract}
Dysarthric speech reconstruction (DSR) typically relies on linguistic or phonetic representations extracted from impaired speech to generate a more intelligible waveform. We investigate a complementary approach that instead intervenes in a representation related to speech production. Starting from a neural analysis--synthesis framework, we introduce a residual mapper that modifies an articulatory-aligned latent space while preserving speaker and prosodic information. The mapper is pretrained on parallel synthetic healthy and artificially dysarthric speech, then adapted to natural dysarthric speech using phoneme-guided and adversarial objectives. We further compare the articulatory bottleneck with a dimension-matched unsupervised representation to assess the benefit of explicit articulatory supervision. 
Objective and perceptual evaluations show that the proposed approach successfully improves artificially generated dysarthric speech. While the transfer to natural pathological speech remains limited, the articulatory bottleneck introduces an interpretable representation, offering a structured space for targeted speech corrections.
\end{abstract}
%

\section{INTRODUCTION}
\label{sec:intro}
Dysarthria is a motor speech disorder in which impaired control of the speech musculature affects respiration, phonation, resonance, articulation, and prosody, often resulting in reduced speech intelligibility \cite{darley1969dysarthria,nishio2006}. Dysarthric speech reconstruction (DSR) aims to transform impaired speech into a more intelligible form while preserving the speaker’s identity \cite{bhat2025}. Recent DSR systems \cite{wang2024unitdsr,chen2024,chen2025diff} typically combine self-supervised learning (SSL) representations of dysarthric speech with explicit speaker information and neural speech generation. While these approaches have shown promising improvements, reconstructing intelligible speech, particularly for severe dysarthria, remains challenging. These systems largely approach DSR as a speech reconstruction problem, in which linguistic or phonetic content is recovered from the impaired signal and used, together with speaker information, to generate a new waveform. Here, we explore a complementary perspective: rather than reconstructing speech primarily from its recovered content, we seek to correct the impaired signal within a representation related to the speech production process. We focus in particular on articulatory movements, a natural intervention space for dysarthria, where impaired motor control can result in inaccurate or poorly coordinated movements of the speech articulators \cite{darley1969dysarthria,nishio2006}.

Our approach builds on two lines of research. First, probing studies have shown that intermediate representations of SSL speech models encode substantial information about speech production and can be strongly correlated with articulatory trajectories \cite{Cho2023,cho2024}. Second, neural speech analysis--synthesis models have exploited such representations to introduce an explicit articulatory bottleneck between speech encoding and waveform reconstruction \cite{georges2024decode,lavechin-hueber-2025-perception,cho2024coding}. Following this approach, we consider an analysis--synthesis framework based on intermediate WavLM representations selected for their strong correspondence with articulatory data, combined with separate prosody and speaker representations and a HiFi-GAN neural vocoder for waveform reconstruction \cite{cho2024coding,kong2020hifi}.

We investigate two alternative bottlenecks derived from the same WavLM representations: a supervised articulatory bottleneck trained to predict articulatory trajectories, and a dimension-matched unsupervised bottleneck obtained using a learned dimensionality-reduction projection. In both cases, a lightweight residual mapper modifies only this compact representation, while the speaker and prosodic representations are preserved. The mapper is first pretrained using parallel synthetic healthy and dysarthric speech and then fine-tuned on natural dysarthric speech using a phoneme-guided CTC objective combined with adversarial regularization toward representations observed in healthy speech.

Our main contributions are:
(1) Formulating dysarthric speech correction as residual mapping within a compact speech-production-related representation while preserving encoded speaker and prosodic information;
(2) Introducing a two-stage training strategy combining supervised mapping from parallel synthetic speech with phoneme-guided and adversarial fine-tuning on natural dysarthric speech;
(3) Comparing supervised articulatory and dimension-matched unsupervised bottlenecks derived from the same SSL representations to investigate the benefit of explicit articulatory supervision.

\begin{figure*}[t]
    \centering
     
    \resizebox{0.7\textwidth}{!}{\input{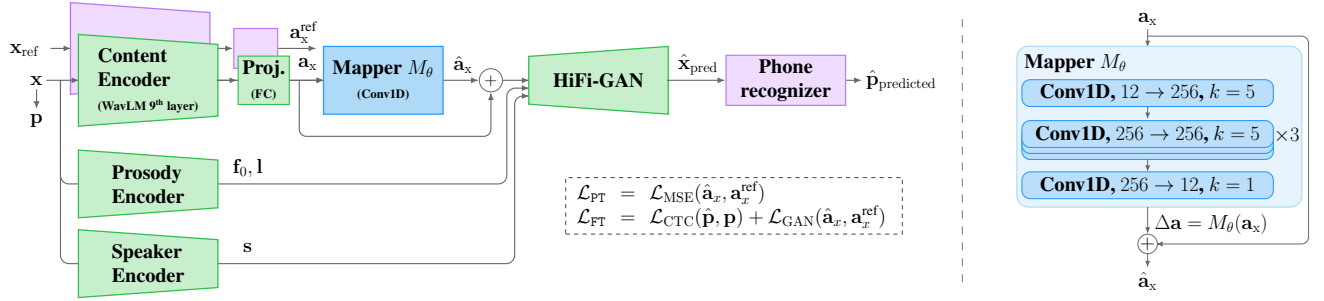}}
    \hspace{0.01\textwidth}
    \tikz[baseline=-0.5ex]{
        \draw[dashed, gray] (0,0) -- (0,3.5cm);
    }
    \hspace{0.01\textwidth}
    \resizebox{0.24\textwidth}{!}{

\definecolor{myorange}{RGB}{255,150,55}
\definecolor{myorangefill}{RGB}{255,224,194}

\definecolor{myblue}{RGB}{40,154,255}
\definecolor{mybluefill}{RGB}{187,222,250}

\definecolor{myblueouter}{RGB}{165,215,250}
\definecolor{myblueouterfill}{RGB}{232,244,253}

\definecolor{mygray}{RGB}{105,105,105}


\begin{tikzpicture}[
    font=\rmfamily,
    >=Latex
]

\path[
    use as bounding box
]
(1,-8) rectangle (10,0.1);


\coordinate (mainAxis) at (5.2,0);

\coordinate (eEMAPos) at (5.2,0);

\coordinate (pseudoPos) at (5.2,-0.5);

\coordinate (mapperOuterPos) at (5.55,-3.2);

\coordinate (conv1Pos) at (5.2,-2.2);
\coordinate (conv2Pos) at (5.2,-3.2);
\coordinate (conv3Pos) at (5.2,-4.95);

\coordinate (addPos) at (5.2,-6.7);

\coordinate (correctedPos) at (5.2,-7.8);

\coordinate (feedbackRight) at (10,0.0);
\coordinate (feedbackDown) at (10,-4.5);

\coordinate (deltaPos) at (7.1,-6.1);


\def\pseudoW{6.2}
\def\pseudoH{0.7}

\def\outerW{8.5}
\def\outerH{4.8}

\def\convW{7.5}
\def\convH{0.8}

\def\correctedW{7.6}
\def\correctedH{0.8}


\node[
    font=\bfseries\LARGE,
]
at (eEMAPos)
{$\mathbf{a}_{\mathrm{x}}$};

\node[
font=\bfseries\LARGE,
]
at (deltaPos)
{$\Delta\mathbf{a}
=
M_{\theta}
(\mathbf{a}_{\mathrm{x}})$};

\node[
    font=\bfseries\LARGE,
]
at (correctedPos)
{$\hat{\mathbf{a}}_{\mathrm{x}}$};


%
%
%




\draw[
    -{Latex[length=2mm]},
    draw=mygray,
    line width=0.8pt
]
let
    \p1 = (eEMAPos),
    \p2 = (mapperOuterPos)
in
    (\x1,\y1-0.3*28)
    --
    (\x1,\y2+\outerH/2*28);


\draw[
    draw=myblueouter,
    fill=myblueouterfill,
    line width=0.9pt,
    rounded corners=10pt
]
($(mapperOuterPos)+(-\outerW/2,\outerH/2)$)
--
($(mapperOuterPos)+(\outerW/2,\outerH/2)$)
--
($(mapperOuterPos)+(\outerW/2,-\outerH/2)$)
--
($(mapperOuterPos)+(-\outerW/2,-\outerH/2)$)
--
cycle;


\draw[
    -{Latex[length=2mm]},
    draw=mygray,
    line width=0.8pt,
    rounded corners=6pt
]
let
    \p1 = (pseudoPos),
    \p2 = (feedbackRight),
    \p3 = (feedbackDown),
    \p4 = (addPos)
in
    (\x1,\y1)
    --
    (\x2,\y1)
    --
    (\x3,\y4)
    --
    (\x4+0.25*28,\y4);


\node[
    font=\bfseries\LARGE,
    align=left
]
at ($(mapperOuterPos)+(-2.5,1.9)$)
{Mapper $M_{\theta}$};


\draw[
    draw=myblue,
    fill=mybluefill,
    line width=0.9pt,
    rounded corners=7pt
]
($(conv1Pos)+(-\convW/2,\convH/2)$)
--
($(conv1Pos)+(\convW/2,\convH/2)$)
--
($(conv1Pos)+(\convW/2,-\convH/2)$)
--
($(conv1Pos)+(-\convW/2,-\convH/2)$)
--
cycle;

\node[
    font=\bfseries\LARGE,
    align=center
]
at (conv1Pos)
{Conv1D, $12 \rightarrow 256$, $k=5$};


\draw[
    -{Latex[length=2mm]},
    draw=mygray,
    line width=0.8pt
]
let
    \p1 = (conv1Pos),
    \p2 = (conv2Pos)
in
    (\x1,\y1-\convH/2*28)--
    (\x2,\y2+\convH/2*28-5);


\draw[
    draw=myblue,
    fill=mybluefill,
    line width=0.9pt,
    rounded corners=7pt
]
($(conv2Pos)+(-\convW/2,-0.18)$)
--
($(conv2Pos)+(\convW/2,-0.18)$)
--
($(conv2Pos)+(\convW/2,-0.18-\convH)$)
--
($(conv2Pos)+(-\convW/2,-0.18-\convH)$)
--
cycle;

\draw[
    draw=myblue,
    fill=mybluefill,
    line width=0.9pt,
    rounded corners=7pt
]
($(conv2Pos)+(-\convW/2,0.00)$)
--
($(conv2Pos)+(\convW/2,0.00)$)
--
($(conv2Pos)+(\convW/2,-\convH)$)
--
($(conv2Pos)+(-\convW/2,-\convH)$)
--
cycle;

\draw[
    draw=myblue,
    fill=mybluefill,
    line width=0.9pt,
    rounded corners=7pt
]
($(conv2Pos)+(-\convW/2,0.18)$)
--
($(conv2Pos)+(\convW/2,0.18)$)
--
($(conv2Pos)+(\convW/2,-\convH+0.18)$)
--
($(conv2Pos)+(-\convW/2,-\convH+0.18)$)
--
cycle;

\node[
    font=\bfseries\LARGE,
    align=center
]
at ($(conv2Pos)+(0,-0.25)$)
{Conv1D, $256 \rightarrow 256$, $k=5$};

\node[
    font=\bfseries\LARGE,
    align=center
]
at ($(conv2Pos)+(\convW/2+0.4,-0.25)$)
{$\times 3$};


\draw[
    -{Latex[length=2mm]},
    draw=mygray,
    line width=0.8pt
]
let
    \p1 = (conv2Pos),
    \p2 = (conv3Pos)
in
    (\x1,\y1-\convH/2*28-0.6*28)
    --
    (\x2,\y2+\convH/2*28);


\draw[
    draw=myblue,
    fill=mybluefill,
    line width=0.9pt,
    rounded corners=7pt
]
($(conv3Pos)+(-\convW/2,\convH/2)$)
--
($(conv3Pos)+(\convW/2,\convH/2)$)
--
($(conv3Pos)+(\convW/2,-\convH/2)$)
--
($(conv3Pos)+(-\convW/2,-\convH/2)$)
--
cycle;

\node[
    font=\bfseries\LARGE,
    align=center
]
at (conv3Pos)
{Conv1D, $256 \rightarrow 12$, $k=1$};


\draw[
    -{Latex[length=2mm]},
    draw=mygray,
    line width=0.8pt
]
let
    \p1 = (mapperOuterPos),
    \p2 = (addPos)
in
    (\x2,\y1-\outerH/2*28-1)
    --
    (\x2,\y2+0.3*28);


\node[
    draw=mygray,
    fill=white,
    circle,
    minimum size=0.58cm,
    inner sep=0pt,
    font=\LARGE
]
(plus)
at (addPos)
{$+$};


%
%


\draw[
    -{Latex[length=2mm]},
    draw=mygray,
    line width=0.8pt
]
let
    \p1 = (addPos),
    \p2 = (correctedPos)
in
    (\x1,\y1-0.30*28)
    --
    (\x2,\y2+\correctedH/2*28);

\end{tikzpicture}}
    \caption{\textbf{Left:} Proposed articulatory correction framework. Frozen encoders extract
    the articulatory bottleneck trajectory $\mathbf{a}_{\mathrm{x}}$, prosodic features (pitch $\mathbf{f}_0$ and loudness $\mathbf{l}$ ), and speaker embedding $\mathbf{s}$ from the input waveform $\mathbf{x}$. 
    A trainable residual Conv1D mapper
    predicts a correction to $\mathbf{a}_{\mathrm{x}}$, yielding
    $\hat{\mathbf{a}}_{\mathrm{x}}$, which is resynthesized by a frozen HiFi-GAN vocoder while preserving the original speaker and prosodic features. 
    Purple boxes are used for training only. A reference audio $\mathbf{x}_\text{ref}$ is sent to the same articulatory bottleneck to pretrain the mapper with MSE loss.
    A frozen phonetic decoder provides phonetic CTC-based supervision against the target sequence $\mathbf{p}$.
    Fine-tuning the mapper combines CTC and adversarial losses. 
    \textbf{Right:} Architecture of the mapper.}
    \label{fig:workflow}
     \vspace{-0.4cm}
\end{figure*}

\section{Related work}
\label{sec:related_work}

Early neural-based DSR systems follow voice conversion paradigms, learning direct mappings between dysarthric and healthy speech representations, such as vocoded features or phonetic posteriorgrams \cite{imai2020improving,purohit2020intelligibility,Zheng2022}. End-to-end encoder--decoder architectures decomposed DSR into sub-modules dedicated to linguistic content extraction, speaker identity preservation, and speech generation, often employing auxiliary objectives such as cross-modal knowledge distillation \cite{wang2020end}, adversarial speaker adaptation \cite{wang2022speaker}, and joint speech reconstruction and phoneme recognition objectives \cite{biadsy2019parrotron,chen21w_interspeech,doshi2021extending}. More recently, SSL-based DSR has emerged as a dominant paradigm, leveraging speech representations extracted from SSL encoders together with neural speech generators, as in Unit-DSR \cite{wang2024unitdsr}, CoLM-DSR \cite{chen2024}, and DiffDSR \cite{chen2025diff}. Despite their architectural differences, these approaches share a common emphasis on recovering linguistic or phonetic information from dysarthric speech to guide waveform reconstruction. In parallel, neural speech analysis--synthesis systems have introduced explicit articulatory bottlenecks inferred from acoustic speech representations \cite{georges2024decode,lavechin-hueber-2025-perception,cho2024coding}. Articulatory features have also been exploited for disordered-speech ASR, including through cross-domain acoustic-to-articulatory inversion \cite{Hu2022,Maharana2021} and data augmentation with articulatory-based synthesis \cite{Xiong2018,Yilmaz2019}. However, these approaches use articulation either for speech analysis--synthesis of typical speech or to help recognition of disordered speech, rather than as an intervention space for DSR. To the best of our knowledge, this work is the first to investigate dysarthric speech reconstruction by explicitly modifying an articulatory representation within a neural speech analysis--synthesis framework.



\section{METHODOLOGY}
\label{sec:methods}

\subsection{Analysis-synthesis pipeline (identity)}

Given an input dysarthric waveform $\mathbf{x}$ and its target phoneme sequence
$\mathbf{p}$, our objective is to generate a waveform $\hat{\mathbf{x}}$ with improved
phonetic intelligibility while preserving the speaker identity and prosodic characteristics\footnote{Dysarthria can also affect prosody and phonation, speech rate, stress patterns, and phoneme voicing. Here, we focus on articulatory configuration.}  of the original signal.

\noindent\textbf{Supervised articulatory bottleneck (\sparc):}
The first proposed model is built on top of SPARC pipeline~\cite{cho2024coding} represented in green in Fig.~\ref{fig:workflow}. 
%
%
%
A content encoder uses the ninth hidden layer of a
pretrained WavLM model~\cite{Chen2022} followed by a projection layer to get a 12-dimensional articulatory bottleneck (AB) $\mathbf{a}_\mathrm{x}$. In SPARC, $\mathbf{a}_\mathrm{x}$ is supervised (\sparc) with an MSE loss to encode the estimated horizontal and vertical trajectories of six EMA sensors attached to the articulators (tongue, lips, jaw, see \cite{richmond11_interspeech}) of a native speaker of English.
In parallel, a frozen prosody encoder extracts the fundamental frequency
$\mathbf{f}_{0}$ and loudness $\mathbf{l}$, and speaker encoder extracts an embedding $\mathbf{s}$
from the input waveform $\mathbf{x}$. A frozen HiFi-GAN-based decoder reconstructs
the waveform from these representations.
When the AB remains unchanged, this reconstruction is
referred to as the \sparc \emph{identity} condition.

\noindent\textbf{Unsupervised articulatory bottleneck (\base):}
To assess whether the correction performance depends on the
pseudo-EMA nature of the \sparc, we construct
a dimension-matched unsupervised articulatory bottleneck (\base). 
The \base model shares exactly the same encoder and decoder as the \sparc model (in green in Fig.~\ref{fig:workflow}), matching its representation dimensionality, while replacing the supervised AB with an unsupervised latent code, using a projection layer consisting of a three-layer MLP with dimensions $1024\!\rightarrow\!512\!\rightarrow\!512\!\rightarrow\!12$ and GELU activations. 
As for \sparc, the projection is applied to the ninth layer of a pretrained WavLM, which linear probing experiments in \cite{cho2024coding} showed to be strongly correlated with EMA trajectories of a target speaker ($r=0.878\pm0.012$). Thus, while \base is not explicitly supervised to predict articulatory movements, its input representation already carries substantial information about speech production.


\subsection{Articulatory mapper}

The mapper $M_{\theta}$ shown in Fig.~\ref{fig:workflow} (right) predicts a residual correction from the
original bottleneck: $\hat{\mathbf{a}}_{\mathrm{x}}~=~\mathbf{a}_{\mathrm{x}}~+~M_{\theta}\left(\mathbf{a}_{\mathrm{x}}\right)$, 
to encourage the mapper to preserve the original
bottleneck when no correction is required.\footnote{\scriptsize{The code and model checkpoints can be found at \url{https://gricad-gitlab.univ-grenoble-alpes.fr/huebert/artimap}}}  The mapper consists
of one-dimensional convolution with 256 channels, followed by three
one-dimensional convolutional blocks with 256 channels and a final
one-dimensional convolution back to the original 12-dimensional bottleneck. 
All intermediate convolutions use a kernel size of~5.

\subsection{Training objectives}


\noindent\textbf{Stage 0, identity models:} The \sparc \textit{identity} model uses the weights provided by SPARC\footnote{\scriptsize{\url{https://github.com/Berkeley-Speech-Group/Speech-Articulatory-Coding}}}~\cite{cho2024coding}. The \base \textit{identity} model is constructed by jointly pretraining the projection with the HiFi-GAN decoder, using the same frozen content, prosody and speaker encoder as in SPARC to provide the input, and the same waveform reconstruction objectives and discriminators as HiFi-GAN training~\cite{kong2020hifi}. 
In the following, and for both \sparc and \base models, all encoders, projection and decoder (green in Fig.~\ref{fig:workflow}) remain frozen throughout training stages 1 and 2 of the mapper, described below.

\noindent\textbf{Stage 1, mapper pre-training:} We use parallel signals of clean and dysarthric synthetic speech ($\mathbf{x}^{\text{ref}}$ and $\mathbf{x}$) (detailed in sec.~\ref{sec:data}) and get their respective AB ($\mathbf{a}^{\text{ref}}_\mathrm{x}$ and $\mathbf{a}_\mathrm{x}$) using a shared frozen content encoder and projection. The mapper is pretrained in a supervised manner so that its output matches $\mathbf{a}^{\text{ref}}_\mathrm{x}$: $
    \mathcal{L}_{\pt}
    =
    \operatorname{MSE}(\mathbf{a}^{\text{ref}}_{\mathrm{x}},\hat{\mathbf{a}}_{\mathrm{x}}).
$
This pre-training exposes the mapper to a wide range of controlled articulatory perturbations, while paired reference trajectories provide direct supervision toward their healthy counterparts.

\noindent\textbf{Stage 2, mapper fine-tuning:} 
The synthesized waveform $\mathbf{\hat{x}}$ is passed through a frozen phone recognizer based on Wav2Vec~2.0\footnote{\scriptsize{\url{https://hf.co/facebook/wav2vec2-lv-60-espeak-cv-ft}}} $R$, producing the predicted phoneme sequence
$\hat{\mathbf{p}}=R(\mathbf{\hat{x}})$. Given the target phoneme sequence $\mathbf{p}$,
the mapper is trained using the combined objective: $
    \mathcal{L}_{\ft}
    =
    \operatorname{CTC}
    \left(R(\mathbf{\hat{x}}),\mathbf{p}\right)
    +
    \lambda_{\mathrm{GAN}}
    \mathcal{L}_{\mathrm{GAN}}(\mathbf{\hat{\mathbf{a}}_{\mathrm{x}}},\mathbf{\hat{\mathbf{a}}^{\text{ref}}_{\mathrm{x}}}).
$
The CTC term promotes the target phonetic sequence, while the adversarial term $\lambda_{\mathrm{GAN}}$ encourages the modified representation $\hat{\mathbf{a}}_{\mathrm{x}}$ to remain within the distribution of healthy speech plausible articulatory trajectories (resp. latent representation). Its associated discriminator $D_{\phi}$ 
contains three one-dimensional convolutional layers with 256 hidden channels, uses a temporal context of two frames, and is trained with label smoothing of $0.1$.
Although the phonetic decoder and HiFi-GAN decoder remain frozen, gradients are propagated through both models to optimize the mapper.

\section{EXPERIMENTS}
\label{sec:experiments}

\subsection{Training data}
\label{sec:data}

\noindent\textbf{Controlled synthetic speech (\Dsynth):}
To pretrain the mapper, we generate parallel clean and artificial dysarthric speech from single-speaker TTS utterances\footnote{\scriptsize{\url{https://hf.co/espnet/kan-bayashi_ljspeech_joint_train_conformer_fastspeech2_hifigan}}}. Each clean utterance is analyzed using the frozen articulatory encoder of SAB to obtain a 12-dimensional articulatory trajectory, which is then selectively perturbed to simulate controlled articulatory deviations. To define these perturbations in an interpretable articulatory space, the trajectories are   converted into six articulatory features following the procedure described in \cite{georges2024decode}, which 
sequentially removes jaw contributions
from the tongue and lip trajectories.
For each articulatory dimension $d$ at time $t$, trajectory excursions were modified
around the corpus mean according to: $\tilde{\mathbf{a}}^{(d)}_{t}~=~\mu_d+\alpha_d(\mathbf{a}^{(d)}_{t}-\mu_d),$ 
where $\alpha_d=1$ preserves the original trajectory and
$0<\alpha_d<1$ compresses its excursion, producing hypo-articulated speech.
The scaling factor $\alpha_d$ was either applied to all six articulatory
features simultaneously or restricted to specific articulatory subsystems.
We used $\alpha_d=0.6$ for moderate
perturbations and $\alpha_d=0.3$ for severe perturbations on the selected
articulatory dimensions. The resulting trajectories were converted back to speech using  the remaining components of the SAB analysis-synthesis pipeline.
We generated \num{5000} clean sentences (\Dsynthclean) and {\num{10000}} perturbed sentences (\Dsynthdys) leading to a total of \SI{7.46}{\hour} of speech. The synthetic test set comprises \num{1000} utterances and \num{28298} reference phonemes, with \num{500} utterances per perturbation condition (moderate and severe).
\newline


\noindent\textbf{Natural dysarthric speech (\Duaspeech):} 
We use six dysarthric speakers from UASpeech \cite{kim2008}: M05, F04, M09, M08, M10, and F05, whose reported intelligibility scores range from \SI{58}{\%} to \SI{95}{\%}, and
with a phoneme deletion rate below \num{0.20}, thereby excluding speakers for whom phoneme omissions are frequent. This selection reflects the scope of our approach: the proposed mapper is designed to correct the articulatory realization of existing phonetic content, rather than recover phonemes that are absent from the input speech.
After trimming the data with voice activity detection and manual inspection, we obtained a dataset of 4424 single-word recordings. To obtain longer inputs while retaining control over their lexical content, we construct three-word dysarthric sequences by concatenating isolated-word recordings from the same speaker. We then constructed training, validation and test set following a $70\%, 15\%, 15\%$ split while ensuring that no lexical item occurs in more than one split. 
The resulting test set comprises \num{224} three-word sequences from the six selected speakers, containing \num{2897} reference phonemes.




\subsection{Experimental setup}

\sparc and \base mappers were both pretrained (stage 1) on the \Dsynth corpus (models $\sparc^{\pt}$ and $\base^{\pt}$). Then, they were fine-tuned (stage 2) either on \Dsynth (models $\sparc^{\ft}_{\synth}$ and $\base^{\ft}_{\synth}$) or \Duaspeech (models $\sparc^{\ft}_{\dys}$ and $\base^{\ft}_{\dys}$) corpus.
We also include for evaluation two models whose mappers were trained with stage 2 only on \Duaspeech ($\base_{\dys}$ and $\sparc_{\dys}$), i.e., without pretraining on synthetic speech.
At evaluation time, we distinguish between the \emph{identity} condition, obtained by analysis--resynthesis without applying the mapper, and the \emph{enhanced} condition, obtained after applying the learned residual transformation. Models trained on \Dsynth are evaluated on both \Dsynth and \Duaspeech test sets, with the latter considered zero-shot evaluation. Models fine-tuned on \Duaspeech are evaluated only on this test set.





\noindent\textbf{Training identity UAB (stage 0):} The \base \textit{identity} projection and decoder are jointly pretrained on
LibriTTS-R for 1.5M steps, using an initial 10k-step
warm-up with an $\ell_1$ log-mel reconstruction loss, followed by
the HiFi-GAN objective combining mel reconstruction, adversarial losses, and feature matching, with respective weights
of \num{45}, \num{1}, and \num{2}. The model is trained using AdamW with a batch size
of \num{16} and one-second waveform segments. 

\noindent\textbf{Pre-training \sparc and \base mappers (stage 1):} Each mapper minimizes MSE over valid frames between mapped pathological codes and frame-aligned healthy targets from \Dsynthdys and \Dsynthclean, respectively, extracted using the corresponding frozen identity encoder and projection. Training uses Adam (batch size 32), a cosine learning-rate schedule from $10^{-3}$ to $10^{-5}$, and gradient clipping at norm 5. We train for up to 50 epochs with early stopping after 10 epochs without validation MSE improvement. 

\noindent\textbf{Fine-tuning \sparc and \base mappers (stage 2):} \sparc and \base mappers are trained using Adam with a learning rate of $10^{-4}$ and a batch
size of 4. Their discriminators are optimized separately with a learning rate
of $10^{-4}$. Fine-tuning runs for up to
\num{120} epochs and validation CTC loss is evaluated every 2 epochs and training stops after 10 evaluations without improvement greater than \num{0.002}. The checkpoint with the lowest validation loss is selected. The adversarial term is weighted by $\lambda_{\mathrm{GAN}}=0.05$.

\subsection{Evaluation metrics}

\noindent\textbf{Automatic metrics:} Phonetic intelligibility is assessed using phoneme error rate (PER). 
Phoneme sequences are predicted with the same phoneme recognizer as in stage 2 training.
We additionally report corpus-level word error rate (WER), transcribing with Whisper large-v3\footnote{\scriptsize{\url{https://hf.co/openai/whisper-large-v3}}} using greedy decoding (English, transcription task, no timestamps).
References and hypotheses are normalized using the Whisper English text normalizer.
\newline

\noindent\textbf{Perceptual evaluation:} 
We conducted an online intelligibility experiment with
\num{48} native English-speaking participants recruited through Prolific.
The experiment comprised three sections. First, participants transcribed
\num{10} synthetic sentences (five moderate and five severe) presented in
four conditions: Healthy TTS, synthetic dysarthric speech, and enhancement
with the unsupervised and speech-articulatory bottlenecks
($\base^{\ft}_{\synth}$ and $\sparc^{\ft}_{\synth}$). Condition assignments
for the synthetic sentences followed a Latin-square rotation across
participants, while sentence order was randomized. Second, participants
transcribed \num{15} isolated words from natural pathological speech, each
presented in one of three conditions: original, \base enhancement
($\base^{\ft}_{\dys}$), or \sparc enhancement ($\sparc^{\ft}_{\dys}$).
Condition assignment for these words was likewise rotated across
participants. Finally, the same \num{15} words were presented in all three
conditions with their corresponding target text, with their order randomized, and participants ranked the
recordings by intelligibility. Transcription intelligibility was measured using WER. Participants
were excluded if their WER on the intelligible TTS controls exceeded 25\%. Audio
examples are available on the accompanying demonstration page.\footnote{\scriptsize{\url{https://www.ultraspeech.com/artimap/demo.html}}}

\begin{table}[t]
    \noindent
        \footnotesize

        \caption{Corpus-level PER and Whisper WER (\%) for SAB and UAB on \Dsynth and \Duaspeech test sets. Identity denotes resynthesis without the mapper; other model rows report enhanced outputs. $\pm$ denotes the half-width of the 95\% CI; $^{\mathrm{ZS}}$ denotes zero-shot evaluation.}
        \vspace{0.1cm}
        \label{tab:main_results}

        \setlength{\tabcolsep}{4pt}

        \begin{tabular}{@{}lrrrr@{}}
            \toprule
            & \multicolumn{2}{c}{\Dsynth test set}
            & \multicolumn{2}{c}{\Duaspeech test set} \\
            \cmidrule(lr){2-3}
            \cmidrule(lr){4-5}

            Model
            & \multicolumn{1}{c}{PER}
            & \multicolumn{1}{c}{WER}
            & \multicolumn{1}{c}{PER}
            & \multicolumn{1}{c}{WER} \\
            \midrule

            Healthy
            & $6.9 \pm 0.4$
            & $4.8 \pm 1.0$
            & \multicolumn{1}{c}{--}
            & \multicolumn{1}{c}{--} \\

            Dysarthric
            & $32.0 \pm 1.4$
            & $35.5 \pm 3.0$
            & ${31.3 \pm 2.8}^{\phantom{\mathrm{ZS}}}$
            & ${44.1 \pm 18.6}^{\phantom{\mathrm{ZS}}}$ \\

            \midrule 
            $\base$ Identity
            & $35.0 \pm 1.4$
            & $38.5 \pm 2.3$
            & ${39.5 \pm 3.5}^{\phantom{\mathrm{ZS}}}$
            & ${53.4 \pm 7.3}^{\phantom{\mathrm{ZS}}}$ \\

            $\base^{\pt}$
            & $12.0 \pm 0.6$
            & $10.6 \pm 1.0$
            & ${77.8 \pm 4.5}^{\mathrm{ZS}}$
            & ${125.0 \pm 9.6}^{\mathrm{ZS}}$ \\

            $\base^{\ft}_{\synth}$
            & $9.4 \pm 0.6$
            & $9.8 \pm 1.0$
            & ${73.3 \pm 4.2}^{\mathrm{ZS}}$
            & ${125.4 \pm 10.0}^{\mathrm{ZS}}$ \\

            $\base^{\ft}_{\dys}$
            & \multicolumn{1}{c}{--}
            & \multicolumn{1}{c}{--}
            & ${34.5 \pm 3.1}^{\phantom{\mathrm{ZS}}}$
            & ${49.2 \pm 5.9}^{\phantom{\mathrm{ZS}}}$ \\

           $\base_{\dys}$
            & \multicolumn{1}{c}{--}
            & \multicolumn{1}{c}{--}
            & ${36.9 \pm 3.1}^{\phantom{\mathrm{ZS}}}$
            & ${51.0 \pm 6.3}^{\phantom{\mathrm{ZS}}}$ \\

            \midrule

            $\sparc$ Identity
            & $35.1 \pm 1.4$
            & $41.9 \pm 2.2$
            & ${50.1 \pm 3.8}^{\phantom{\mathrm{ZS}}}$
            & ${76.8 \pm 7.7}^{\phantom{\mathrm{ZS}}}$ \\

            $\sparc^{\pt}$
            & $20.8 \pm 1.0$
            & $23.0 \pm 1.7$
            & ${61.9 \pm 4.6}^{\mathrm{ZS}}$
            & ${133.1 \pm 41.2}^{\mathrm{ZS}}$ \\

            $\sparc^{\ft}_{\synth}$
            & $17.4 \pm 0.9$
            & $19.4 \pm 1.5$
            & ${71.0 \pm 5.1}^{\mathrm{ZS}}$
            & ${125.1 \pm 9.4}^{\mathrm{ZS}}$ \\

            $\sparc^{\ft}_{\dys}$
            & \multicolumn{1}{c}{--}
            & \multicolumn{1}{c}{--}
            & ${38.5 \pm 3.1}^{\phantom{\mathrm{ZS}}}$
            & ${78.9 \pm 18.2}^{\phantom{\mathrm{ZS}}}$ \\

            $\sparc_{\dys}$
            & \multicolumn{1}{c}{--}
            & \multicolumn{1}{c}{--}
            & ${42.6 \pm 3.3}^{\phantom{\mathrm{ZS}}}$
            & ${70.2 \pm 7.7}^{\phantom{\mathrm{ZS}}}$ \\

            \bottomrule

        \end{tabular}

    %
\end{table}

\section{Results}

\noindent\textbf{Bottleneck mapping corrects synthetic perturbations.} As shown in Table 1, MSE pretraining substantially reduces recognition errors. Compared with a \Dsynth PER of \SI{32.0}{\%} for the perturbed input, $\base^{\pt}$ and $\sparc^{\pt}$ achieve \SI{12.0}{\%} and \SI{20.8}{\%}, respectively. Fine-tuning with the combined CTC and adversarial objective brings further improvements, reaching \SI{9.4}{\%} for $\base^{\ft}_{\synth}$ and \SI{17.4}{\%} for $\sparc^{\ft}_{\synth}$. A similar pattern is observed for WER, which decreases from \SI{35.5}{\%} for the input to \SI{10.6}{\%} and \SI{23.0}{\%} after pretraining, and to \SI{9.8}{\%} and \SI{19.4}{\%} after fine-tuning.
The perceptual transcription experiment confirms the benefit of enhancement on synthetic speech: human WER decreased from \SI{29.7}{\%} for the synthetic dysarthric condition to \SI{21.6}{\%} with $\base^{\ft}_{\synth}$. In contrast, $\sparc^{\ft}_{\synth}$ yielded a WER of \SI{31.8}{\%}, indicating that the objective improvements obtained with the articulatory mapper did not translate into a perceptual advantage in this setting. Clean TTS speech reached \SI{5.0}{\%} human WER, providing an upper-bound control for the perceptual evaluation.
\newline

\noindent\textbf{Adaptation to natural speech enables correction within the resynthesis pipeline.}
Zero-shot (ZS) experiments involving models trained on synthetic only speech yield \Duaspeech PER values between \SI{61.9}{\%} and \SI{77.8}{\%}, above their identity baselines, suggesting that controlled perturbations in \Dsynthdys do not fully capture natural dysarthric variability, and that adaptation to natural speech is necessary.
 After fine-tuning on \Duaspeech, PER decreases from \SI{39.5}{\%} to \SI{34.5}{\%} for \base and from \SI{50.1}{\%} to \SI{38.5}{\%} for \sparc relative to their identity conditions. 
The improvements after natural-speech fine-tuning show that the mapper can learn from target-domain data.
Comparison between $\base^{\ft}_{\dys}$ and $\base_{\dys}$ (resp. $\sparc^{\ft}_{\dys}$ and $\sparc_{\dys}$) shows the benefit  on PER of model pretraining on synthetic speech.
However, all enhanced systems remain above the original-input PER of \SI{31.3}{\%}. Word-level results also differ: WER decreases from \SI{53.4}{\%} to \SI{49.2}{\%} for \base, while increasing from \SI{76.8}{\%} to \SI{78.9}{\%} for \sparc, compared with \SI{44.1}{\%} for the original recordings. Surprisingly, the stage 1 pretraining is not beneficial in this case. 
The human transcription experiment showed the same overall tendency on natural pathological speech: human WER was \SI{53.3}{\%} for the original recordings, \SI{55.8}{\%} after \base enhancement, and \SI{62.9}{\%} after SAB enhancement.
In terms of intelligibility ranking, Wilcoxon tests showed
that the original recordings received significantly better ranks than
$\sparc$ ($p=.0026$). No significant differences were observed
between the original recordings and $\base$, or between $\base$ and $\sparc$
($p=.074$ for both comparisons).
\newline



\begin{figure}[t]
    \centering
    \includegraphics[width=0.9\linewidth]{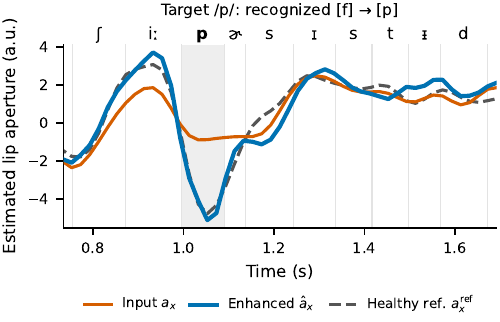}
     \vspace{-0.3cm}
    \caption{Estimated lip aperture for a synthetic test utterance,``She persisted.'' Around the target /p/ (shaded), the trajectory reconstructed by $\sparc^{\ft}_{\synth}$ approaches the healthy reference, accompanying a change in recognition from [f] to [p].}
    \label{fig:EMA}
\end{figure}

\noindent\textbf{Both bottlenecks support correction, while \sparc allows changes to be interpreted in articulatory terms.}
The comparison between \sparc and \base provides insight into the role of explicit articulatory supervision. On synthetic speech, their identity PER values are nearly identical (\SI{35.1}{\%} and \SI{35.0}{\%}), whereas \base achieves lower recognition error after both mapper training stages. \sparc nevertheless offers an interpretable structure: its dimensions correspond to estimated articulator trajectories, allowing the mapper's changes to be inspected. This provides a basis for investigating how bottleneck modifications relate to phonetic correction. Fig.~\ref{fig:EMA} displays an example of the lip aperture extracted from dysarthric ($\mathbf{a}_{\mathrm{x}}$), reconstructed ($\hat{\mathbf{a}}_{\mathrm{x}}$) and healthy reference ($\hat{\mathbf{a}}^{\text{ref}}_{\mathrm{x}}$) AB with the $\sparc^{\ft}_{\synth}$ model. Around the target /p/ (shaded), the trajectory reconstructed by $\sparc^{\ft}_{\synth}$ approaches the healthy reference.


\section{CONCLUSION}


This work introduces the first approach to explicitly modify articulatory representations for dysarthric speech reconstruction within a frozen neural analysis--synthesis framework. We address here the root cause of reduced intelligibility in dysarthria, rather than merely reconstructing linguistic content. We demonstrate an effective curriculum training pipeline: pretraining the mapper with MSE on synthetic speech provides an efficient initialization while fine-tuning with a phoneme-guided CTC objective on dysarthric speech then enables adaptation without requiring parallel healthy references. This pipeline achieves strong performance on synthetic data, validating the feasibility of corrective transformations within a compact bottleneck.
While transferring these improvements to natural dysarthric speech remains challenging, we attribute this primarily to the scarcity of high-quality dysarthric data. Crucially, our method introduces an interpretable bottleneck, where each dimension corresponds to a specific articulatory trajectory. 
We believe that pursuing this direction with newly recorded data will enable precise, localized modifications with significant applications in speech rehabilitation, including visualization and coaching tools for clinical practice.

\newpage
\section{Acknowledgment}
This research was supported by the BPI project THERADIA and by MIAI Cluster (ANR-23-IACL-0006).

\section{Ethical Compliance and AI Use Disclosure}
Speech corpora were used under their respective terms of use. Prolific participants provided informed consent and received compensation. Tools from Anthropic, OpenAI, and Mistral assisted with manuscript editing and code development. The authors reviewed and edited the resulting content and take full responsibility for the publication.
\bibliographystyle{IEEEbib}
\bibliography{strings,refs}

@INPROCEEDINGS{Cho2023,
  author={Cho, Cheol Jun and Wu, Peter and Mohamed, Abdelrahman and Anumanchipalli, Gopala K.},
  booktitle = {\textcolor{white}{ } ICASSP},
  title={Evidence of Vocal Tract Articulation in Self-Supervised Learning of Speech}, 
  year={2023},
  volume={},
  number={},
  pages={1-5},
  doi={10.1109/ICASSP49357.2023.10094711}
}

@article{bhat2025,
  title={Speech technology for automatic recognition and assessment of dysarthric speech: An overview},
  author={Bhat, Chitralekha and Strik, Helmer},
  journal={Journal of Speech, Language, and Hearing Research},
  volume={68},
  number={2},
  pages={547--577},
  year={2025},
  publisher={American Speech-Language-Hearing Association}
}

@inproceedings{kim2008,
  title={Dysarthric speech database for universal access research},
  author={Kim, Heejin and Hasegawa-Johnson, Mark and Perlman, Adrienne and Gunderson, Jon and Huang, Thomas S and Watkin, Kenneth and Frame, Simone},
  booktitle={\textcolor{white}{ } Interspeech},
  pages={1741--1744},
  year={2008}
}

@inproceedings{chen2025diff,
  title     = {{DiffDSR: Dysarthric Speech Reconstruction Using Latent Diffusion Model}},
  author    = {Xueyuan Chen and Dongchao Yang and Wenxuan Wu and Minglin Wu and Jing Xu and Xixin Wu and Zhiyong Wu and Helen Meng},
  year      = {2025},
booktitle = {\textcolor{white}{ } {Interspeech}},
  pages     = {2113--2117},
  doi       = {10.21437/Interspeech.2025-1770},
  issn      = {2958-1796},
}

@inproceedings{cho2024,
  title={Self-supervised models of speech infer universal articulatory kinematics},
  author={Cho, Cheol Jun and Mohamed, Abdelrahman and Black, Alan W and Anumanchipalli, Gopala K},
  booktitle={\textcolor{white}{ } ICASSP},
  pages={12061--12065},
  year={2024}
}

@article{nishio2006,
  title={Comparison of speaking rate, articulation rate and alternating motion rate in dysarthric speakers},
  author={Nishio, Masaki and Niimi, Seiji},
  journal={Folia Phoniatrica et Logopaedica},
  volume={58},
  number={2},
  pages={114--131},
  year={2006},
  publisher={S. Karger AG Basel, Switzerland}
}

@article{cho2024coding,
  title={Coding speech through vocal tract kinematics},
  author={Cho, Cheol Jun and Wu, Peter and Prabhune, Tejas S and Agarwal, Dhruv and Anumanchipalli, Gopala K},
  journal={IEEE Journal of Selected Topics in Signal Processing},
  volume={18},
  number={8},
  pages={1427--1440},
  year={2024},
  publisher={IEEE}
}

@inproceedings{chen2024,
  title     = {{CoLM-DSR: Leveraging Neural Codec Language Modeling for Multi-Modal Dysarthric Speech Reconstruction}},
  author    = {Xueyuan Chen and Dongchao Yang and Dingdong Wang and Xixin Wu and Zhiyong Wu and Helen Meng},
  year      = {2024},
  booktitle = {\textcolor{white}{ } {Interspeech}},
  pages     = {4129--4133},
  doi       = {10.21437/Interspeech.2024-1852},
  issn      = {2958-1796},
}

@article{Chen2022,
	author = {Chen, Sanyuan and Wang, Chengyi and Chen, Zhengyang and Wu, Yu and Liu, Shujie and Chen, Zhuo and Li, Jinyu and Kanda, Naoyuki and Yoshioka, Takuya and Xiao, Xiong and Wu, Jian and Zhou, Long and Ren, Shuo and Qian, Yanmin and Qian, Yao and Wu, Jian and Zeng, Michael and Yu, Xiangzhan and Wei, Furu},
	journal = {IEEE Journal of Selected Topics in Signal Processing},
	number = {6},
	pages = {1505--1518},
	title = {{WavLM}: Large-Scale Self-Supervised Pre-Training for Full Stack Speech Processing},
	volume = {16},
    doi={10.1109/JSTSP.2022.3188113},
	year = {2022}}

@article{kong2020hifi,
  title={{HiFi-GAN}: Generative adversarial networks for efficient and high fidelity speech synthesis},
  author={Kong, Jungil and Kim, Jaehyeon and Bae, Jaekyoung},
  journal={Advances in Neural Information Processing Systems},
  volume={33},
  pages={17022--17033},
  year={2020},
  url={https://dl.acm.org/doi/abs/10.5555/3495724.3497152}
}

@article{darley1969dysarthria,
  author = {Darley, F. L. and Aronson, A. E. and Brown, J. R.},
  title = {Differential Diagnostic Patterns of Dysarthria},
  journal = {Journal of Speech and Hearing Research},
  volume = {12},
  number = {2},
  pages = {246--269},
  year = {1969}
}

@article{georges2024decode,
  author = {Georges, M.-A. and Lavechin, M. and Schwartz, J.-L. and Hueber, T.},
  title = {Decode, Move and Speak! Self-supervised Learning of Speech Units, Gestures, and Sound Relationships Using Vocal Imitation},
  journal = {Computational Linguistics},
  volume = {50},
  number = {4},
  pages = {1345--1373},
  year = {2024}
}

@inproceedings{imai2020improving,
  author = {Imai, S. and Nose, T. and Kanagaki, A. and Watanabe, S. and Ito, A.},
  title = {Improving Pronunciation Clarity of Dysarthric Speech Using {CycleGAN} with Multiple Speakers},
  booktitle = {\textcolor{white}{ } {GCCE}},
  pages = {366--367},
  year = {2020}
}

@inproceedings{purohit2020intelligibility,
  author = {Purohit, M. and others},
  title = {Intelligibility Improvement of Dysarthric Speech Using {MMSE} {DiscoGAN}},
  booktitle = {\textcolor{white}{ } {SPCOM}},
  pages = {1--5},
  year = {2020}
}

@inproceedings{wang2020end,
  author = {Wang, D. and others},
  title = {End-to-End Voice Conversion via Cross-Modal Knowledge Distillation for Dysarthric Speech Reconstruction},
  booktitle = {\textcolor{white}{ } ICASSP},
  pages={7744-7748},
 year = {2020}
}

@inproceedings{wang2022speaker,
  author = {Wang, D. and others},
  title = {Speaker Identity Preservation in Dysarthric Speech Reconstruction by Adversarial Speaker Adaptation},
  booktitle = {\textcolor{white}{ } ICASSP},
  pages={6677-6681},
  year = {2022}
}

@inproceedings{biadsy2019parrotron,
  title     = {{Parrotron: An End-to-End Speech-to-Speech Conversion Model and its Applications to Hearing-Impaired Speech and Speech Separation}},
  author    = {Fadi Biadsy and Ron J. Weiss and Pedro J. Moreno and Dimitri Kanevsky and Ye Jia},
  year      = {2019},
  booktitle = {{\textcolor{white}{ } Interspeech}},
  pages     = {4115--4119},
  doi       = {10.21437/Interspeech.2019-1789},
  issn      = {2958-1796},
}

@inproceedings{chen21w_interspeech,
  title     = {{Conformer Parrotron: A Faster and Stronger End-to-End Speech Conversion and Recognition Model for Atypical Speech}},
  author    = {Zhehuai Chen and Bhuvana Ramabhadran and Fadi Biadsy and Xia Zhang and Youzheng Chen and Liyang Jiang and Fang Chu and Rohan Doshi and Pedro J. Moreno},
  year      = {2021},
  booktitle = {{\textcolor{white}{ } Interspeech}},
  pages     = {4828--4832},
  doi       = {10.21437/Interspeech.2021-676},
}

@inproceedings{doshi2021extending,
  author = {Doshi, R. and others},
  title = {Extending Parrotron: An End-to-End Speech Conversion and Speech Recognition Model for Atypical Speech},
  booktitle = {\textcolor{white}{ } ICASSP},
  pages = {6988-6992},
  year = {2021}
}

@INPROCEEDINGS{wang2024unitdsr,
  author={Wang, Yuejiao and Wu, Xixin and Wang, Disong and Meng, Lingwei and Meng, Helen},
  booktitle={\textcolor{white}{ } {ICASSP}},
  title={{UNIT-DSR}: Dysarthric Speech Reconstruction System Using Speech Unit Normalization},
  year={2024},
  pages={12306--12310},
  doi={10.1109/ICASSP48485.2024.10446921}
}

@INPROCEEDINGS{Hu2022,
  author={Hu, Shujie and Liu, Shansong and Xie, Xurong and Geng, Mengzhe and Wang, Tianzi and Hu, Shoukang and Cui, Mingyu and Liu, Xunying and Meng, Helen},
  booktitle={\textcolor{white}{ } ICASSP}, 
  title={Exploiting Cross Domain Acoustic-to-Articulatory Inverted Features for Disordered Speech Recognition}, 
  year={2022},
  pages={6747-6751},
  doi={10.1109/ICASSP43922.2022.9746989}}

@INPROCEEDINGS{Xiong2018,
  author={Xiong, Feifei and Barker, Jon and Christensen, Heidi},
  booktitle= {\textcolor{white}{ } {ITG-Symposium}}, 
  title={Deep Learning of Articulatory-Based Representations and Applications for Improving Dysarthric Speech Recognition}, 
  year={2018},
  volume={},
  number={},
  pages={1-5},
  doi={}}

@article{Yilmaz2019,
title = {Articulatory and bottleneck features for speaker-independent ASR of dysarthric speech},
journal = {Computer Speech \& Language},
volume = {58},
pages = {319-334},
year = {2019},
doi = {https://doi.org/10.1016/j.csl.2019.05.002},
url = {https://www.sciencedirect.com/science/article/pii/S0885230818302663},
author = {Emre Yılmaz and Vikramjit Mitra and Ganesh Sivaraman and Horacio Franco},
}

@article{Zheng2022,
title = {Phonetic posteriorgram-based voice conversion system to improve speech intelligibility of dysarthric patients},
journal = {Computer Methods and Programs in Biomedicine},
volume = {215},
pages = {106602},
year = {2022},
doi = {https://doi.org/10.1016/j.cmpb.2021.106602},
url = {https://www.sciencedirect.com/science/article/pii/S0169260721006763},
author = {Wei-Zhong Zheng and Ji-Yan Han and Chen-Kai Lee and Yu-Yi Lin and Shu-Han Chang and Ying-Hui Lai},
}

@INPROCEEDINGS{Maharana2021,
  author={Maharana, Sarthak Kumar and Illa, Aravind and Mannem, Renuka and Belur, Yamini and Shetty, Preetie and Kumar, Veeramani Preethish and Vengalil, Seena and Polavarapu, Kiran and Atchayaram, Nalini and Ghosh, Prasanta Kumar},
  booktitle={\textcolor{white}{ } ICASSP}, 
  title={Acoustic-to-Articulatory Inversion for Dysarthric Speech by Using Cross-Corpus Acoustic-Articulatory Data}, 
  year={2021},
  pages={6458-6462},
  doi={10.1109/ICASSP39728.2021.9413625}}

@inproceedings{richmond11_interspeech,
  title     = {{Announcing the electromagnetic articulography (day 1) subset of the mngu0 articulatory corpus}},
  author    = {Korin Richmond and Phil Hoole and Simon King},
  year      = {2011},
  booktitle = {{\textcolor{white}{ } Interspeech}},
  pages     = {1505--1508},
  doi       = {10.21437/Interspeech.2011-316},
}

@inproceedings{lavechin-hueber-2025-perception,
  author={Lavechin, Marvin and Hueber, Thomas},
  title={From perception to production: how acoustic invariance facilitates articulatory learning in a self-supervised vocal imitation model},
  booktitle={\textcolor{white}{ } {EMNLP}},
  year={2025},
  pages={23852--23863},
  doi={10.18653/v1/2025.emnlp-main.1217}
}

\end{document}